\documentstyle[sprocl,epsfig]{article}

\def\Journal#1#2#3#4{{#1} {\bf #2}, #3 (#4)}
\def\NPB{{\em Nucl. Phys.} B}
\def\NPA{{\em Nucl. Phys.} A}
\def\PLB{{\em Phys. Lett.}  B}

\def\PRD{{\em Phys. Rev.} D}

\newcommand{\beq}{\begin{equation}}
\newcommand{\eeq}{\end{equation}}
\newcommand{\la}[1]{\label{#1}}
\newcommand{\ba}{\begin{array}}
\newcommand{\ea}{\end{array}}
\newcommand{\ur}[1]{(\ref{#1})}

\newcommand{\Eq}[1]{Eq.~(\ref{#1})}

 \def\Dirac#1{#1\hskip-5.5pt/}
 \def\dd{\Dirac\partial}

\begin{document}

\title{INSTANTONS AND BARYON DYNAMICS}

\author{DMITRI DIAKONOV}

\address{NORDITA, Blegdamsvej 17, DK-2100 Copenhagen, Denmark\\
E-mail: diakonov@nordita.dk}

\address{PNPI, Gatchina, St. Petersburg 188 300, Russia}

\maketitle

\abstracts{I explain how instantons break chiral symmetry
and how do they bind quarks in baryons. The confining potential
is possibly irrelevant for that task.}

\section{Introduction}

According to common wisdom, moving a quark away from a diquark system
in a baryon generates a string, also called a flux tube, whose energy
rises linearly with the separation. The string energy, however, exceeds
the pion mass $m_\pi=140\,{\rm MeV}$ at a modest separation of about
$0.26\,{\rm fm}$, see Fig. 1. At larger
separations the would-be linear potential is screened since it is
energetically favorable to tear the string and produce a pion. Virtually,
the linear potential can stretch to as much as $0.4\,{\rm fm}$ where its
energy exceeds $2m_\pi$ but that can happen only for a short time of
$1/m_\pi$. Meanwhile, the ground-state baryons are stable,
and their sizes are about $1\,{\rm fm}$. The pion-nucleon coupling
is huge, and there seems to be no suppression of the string breaking
by pions. The paradox is that the linear potential of the pure glue world,
important as it might be to explain why quarks are not observed as a matter of
principle, can hardly have a direct impact on the properties of lightest
hadrons. What, then, determines their structure?

\begin{figure}[t]
\epsfig{file=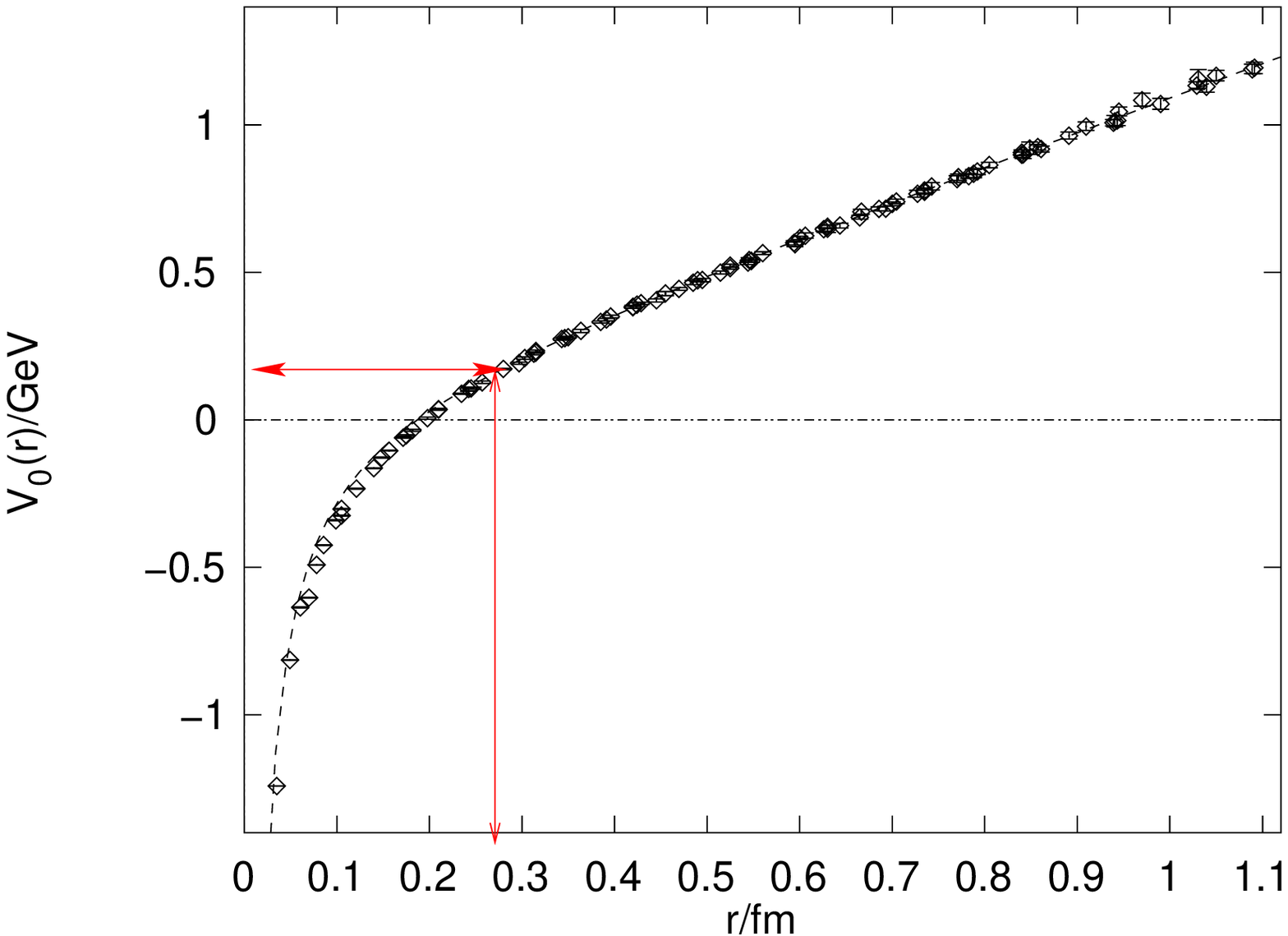,width=5.7cm}
\hspace{-.01cm}\epsfig{file=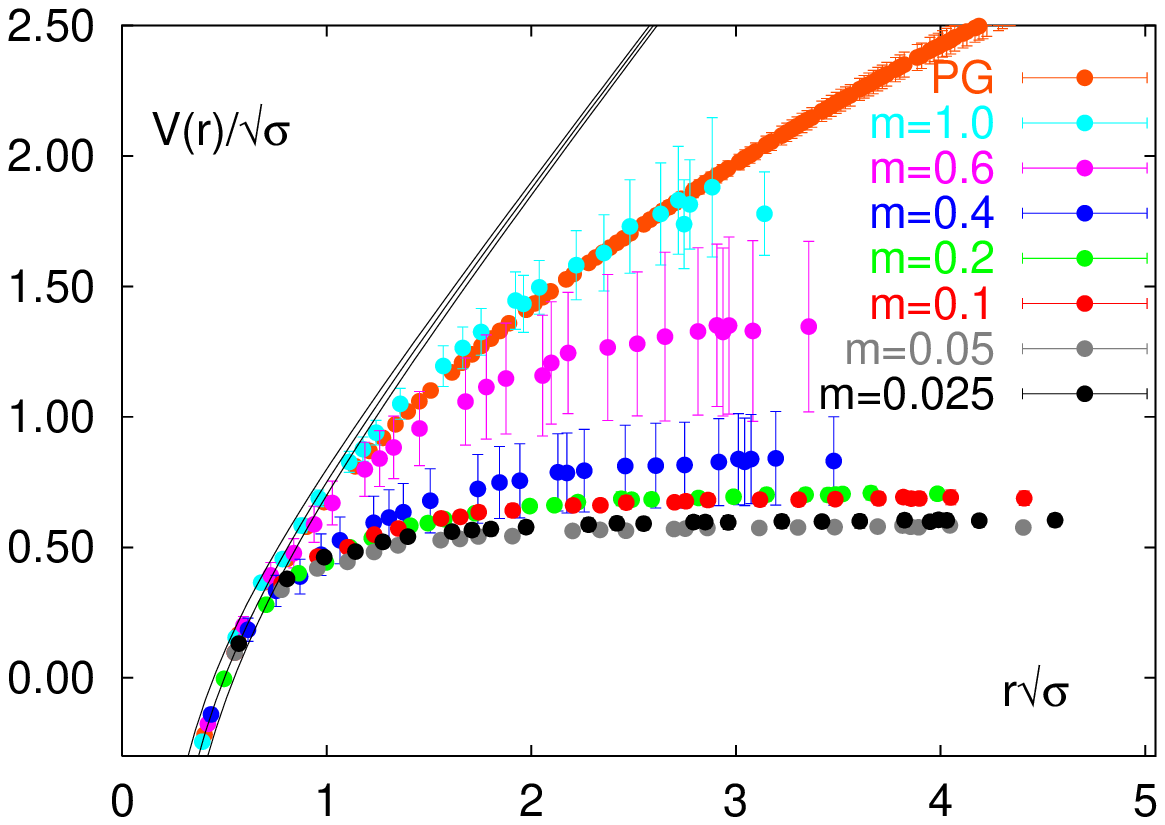,width=6cm}
\vspace{-.01cm}\caption{The lattice-simulated potential between static
quarks $^1$ exceeds $m_\pi$ at the separation of $0.26\,{\rm fm}$ (left).
The screening of the linear potential is clearly seen in simulations
at high temperatures but below the phase transition $^2$ (right).
As one lowers the pion mass the string breaking happens at smaller
distances; the scale is
$\sqrt{\sigma}\simeq 425\,{\rm MeV}\simeq (0.47\,{\rm fm})^{-1}$.
\label{scrpot}}
\end{figure}

We know that, were the chiral symmetry of QCD unbroken, the
lightest hadrons would appear in parity doublets. The large actual
splitting between, say, $N(\frac{1}{2}^-,1535)$ and
$N(\frac{1}{2}^+,940)$
implies that chiral symmetry is spontaneously broken as
characterized by the nonzero quark condensate
$<\!\bar q q\!>\simeq -(250\,{\rm MeV})^3$. Equivalently, it means
that nearly massless (`current') quarks obtain a sizable non-slash
term in the propagator, called the dynamical or constituent mass
$M(p)$, with $M(0)\simeq 350\,{\rm MeV}$. The $\rho$-meson has
roughly twice and nucleon thrice this mass, {\it i.e.} are relatively
loosely bound. The pion is a (pseudo) Goldstone boson and is
very light. The sizes of these hadrons are typically $\sim 1/M(0)$
whereas the size of constituent quarks is given by the slope
of $M(p)$ and is much less. It explains, at least on the qualitative
level, why constituent quark models are so phenomenologically
successful.

We see thus that the spontaneous chiral symmetry breaking (SCSB)
rather than the expected linear confining potential of the pure glue world
is key to the understanding the origin of the ground-state hadrons.
It may be that for highly excited hadrons the importance of confinement
forces {\it vs} SCSB is reversed: I discuss it at the end of the paper.
In the main part I briefly review the instanton mechanism of the SCSB
suggested and worked out by Victor Petrov and myself in the middle of
the 80's. \cite{DP1,DP2} Much analytical and numerical work calculating
hadron observables has supported this mechanism, see Refs.~\cite{DSS,DP3}
for reviews. There is also growing support from direct lattice
simulations, see below. Instantons induce strong interaction between
quarks, leading to bound-state baryons with calculable and reasonable
properties.

\section{What are instantons?}
Being a quantum field theory, QCD deals with the fluctuating gluon and quark
fields. A fundamental fact \cite{FJR} is that the potential energy
of the gluon field is a periodic function in one particular direction
in the infinite-dimensions functional space; in all other directions the
potential energy is oscillator-like. This is illustrated in Fig. 2.

\begin{figure}[t]
\epsfig{file=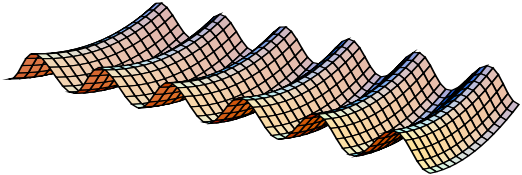,width=7cm,height=3cm}
\vspace{-.01cm}\caption{Potential energy of the gluon field is periodic in one direction
and oscillator-like in all other directions in functional space.
\label{periodic}}
\end{figure}

{\bf Instanton is a large fluctuation of the gluon field} in imaginary
(or Euclidean) time corresponding to quantum tunneling from one minimum
of the potential energy to the neighbor one. Mathematically, it was
discovered by Belavin, Polyakov, Schwarz and Tiupkin; \cite{BPST}
the tunneling interpretation was given by V. Gribov. \cite{Pol}
The name `instanton' has been introduced by 't~Hooft~\cite{tH} who
studied many of the key properties of those fluctuations. Anti-instantons
are similar fluctuations but tunneling in the opposite direction
in Fig. 2.

\begin{figure}[t]
\epsfxsize=25pc % will enlarge or reduce the postscript figures
%based on the xsize
\epsfbox{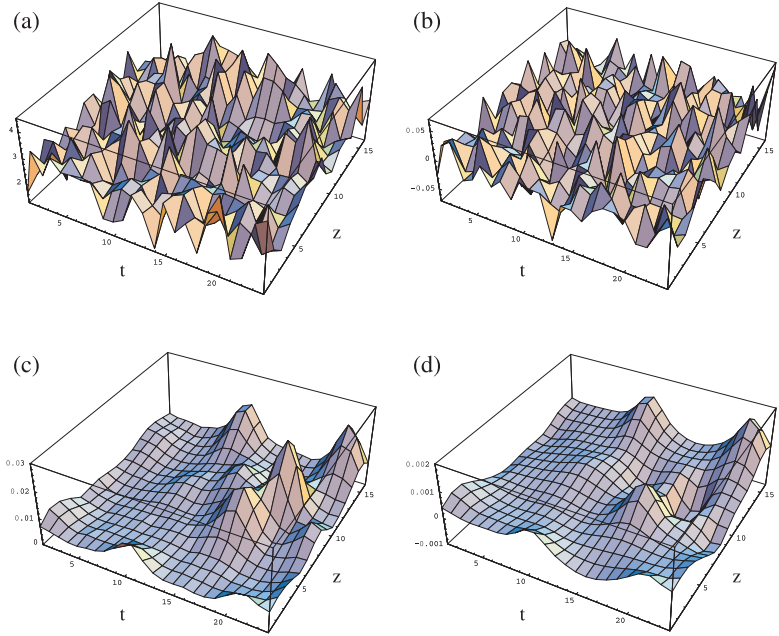} % postscript image file name
\caption{Smoothing out the normal zero-point oscillations
reveals large fluctuations of the gluon field, which are
nothing but instantons and anti-instantons with random positions
and sizes. The left column shows the action density and the right column
shows the topological charge density for the same snapshot. $^{11}$
\label{fig:negele1}}
\end{figure}

Instanton fluctuations are characterized by their position in space-time
$z_\mu$, the spatial size $\rho$ and orientation in color space $O$,
all in all by 12 collective coordinates. The probability for the instanton
fluctuation is, roughly, given by the WKB tunneling amplitude,
\beq
\exp(-{\rm Action})=\exp\left(-\frac{1}{4g^2}\int\!d^4x\,F_{\mu\nu}^2\right)
=\exp\left(-\frac{8\pi^2}{g^2}\right).
\la{ta}\eeq
It is non-analytic in the gauge coupling constant and hence instantons are
missed in all orders of the perturbation theory. However, it is not
a reason to ignore tunneling. For example, tunneling of electrons from one
atom to another in a metal is also a nonperturbative effect but we would get
nowhere in understanding metals had we ignored it. Indeed, instantons are
clearly seen in nonperturbative lattice simulations of the gluon vacuum.
In the upper part of Fig. 3 (taken from the paper by J. Negele {\it et al.}
\cite{N}) a typical snapshot of gluon fluctuations in the vacuum
is shown. Naturally, it is heavily dominated by normal perturbative
UV-divergent zero-point oscillations of the field. However, after
smearing out these oscillations (there are now several techniques
developed how to do it) one reveals a smooth background field which
has proven to be nothing but an ensemble of instantons and anti-instantons
with random positions and sizes. The lower part of Fig. 3 is what is left
of the upper part after smoothing.

The average size of instantons found in ref. \cite{N} is
$\bar\rho\approx 0.36\,{\rm fm}$ and their average separation
is $\bar R=(N/V)^{-\frac{1}{4}}\approx 0.89\,{\rm fm}$. Similar
results have been obtained by other lattice groups using various
techniques. A decade earlier the basic characteristics of the
instanton ensemble were obtained analytically from the Feynman
variational principle \cite{DP4,DPW} and expressed through the only
dimensional parameter $\Lambda$ one has in QCD:
$\bar\rho\approx 0.48/\Lambda_{\overline{\rm MS}}\simeq 0.35\,{\rm fm}$,
$\bar R\approx 1.35/\Lambda_{\overline{\rm MS}}\simeq 0.95\,{\rm fm}$,
if one uses $\Lambda_{\overline{\rm MS}}=280\,{\rm MeV}$ as it follows
from the DIS data.

The theory of the instanton vacuum is based on the
assumption that the QCD `partition function' is saturated by large
nonperturbative fluctuations of the gluon field (instantons), plus
perturbative oscillations about them. It takes the form of a
partition function of a liquid (like ${\rm H_2O}$) of $N_+$ instantons
and $N_-$ anti-instantons:
\beq
{\cal Z}=\sum_{N_\pm}\frac{1}{N_+!N_-!}\prod_{I,\bar I}
\int d^4z\,d^7O\, \frac{d\rho}{\rho^5}\,(\rho\Lambda)^{11}\,
e^{-U_{\rm int}}
\la{Z}\eeq
where $U_{\rm int}$ is the interaction depending on relative positions,
sizes and orientations of instantons.
Actually, it is a {\it grand canonical ensemble} of interacting `particles'
since their numbers $N_\pm$ are not fixed but must be found from the
minimum of the free energy and ultimately expressed through
$\Lambda =\frac{1}{a}\,\exp(-\frac{8\pi^2}{11g^2})$
appearing through the `transmutation of dimensions' from integrating over
perturbative gluons ($a$~is the UV cutoff, e.g. the lattice spacing).
The numbers for $\bar\rho$ and $\bar R$ quoted above come from the study
of this ensemble.~\cite{DP4,DPW}~\footnote{The first study
of the instanton ensemble on a qualitative level was performed by Callan,
Dashen and Gross \cite{CDG} and later on by Shuryak \cite{S}.
Ilgenfritz and Mueller-Preussker \cite{IMP} were the first to study
the instanton ensemble quantitatively by modelling the interactions by
a hard-core repulsion.}

\section{How do instantons break chiral symmetry?}
We now switch in light quarks into the random instanton ensemble.
The basic property is that massless quarks are bound by instantons
with exactly zero `energy'. These localized states are called {\em
quark zero modes}, discovered by 't Hooft \cite{tH}. They have definite
helicity or chirality: left-handed quarks are localized on instantons ($I$)
and right-handed are on anti-instantons ($\bar I$).

However, this is correct only for a single (anti)instanton. If there is a
$I\bar I$ pair, no matter how far apart, the degeneracy of the two would-be
exactly zero  modes is lifted owing to the overlap of their wave functions.
If there are infinitely many $I$'s and $\bar I$'s, each of them brings in
a would-be zero mode but, because of the quantum-mechanical overlap,
the degenerate levels split and form a continuous spectrum,
meaning the {\em delocalization} of the would-be zero modes. \cite{DP1}
The effect is similar to the so-called Anderson conductivity:
the appearance of the conductivity of electrons bound by random impurities.

It can be shown mathematically that a finite density of quark
states at zero `energy' means spontaneous chiral symmetry breaking,
and one can calculate the chiral condensate from the average overlap of the
zero-mode wave functions and express it through the basic quantities
characterizing the instanton ensemble, i.e. the average size of instantons
and their density \cite{DP1}. However, there is a simpler physical argument.
Each time a quark `hops' from one random instanton to another it has to change
its helicity. Delocalization implies quarks make an infinite number of such
jumps. An infinite number of helicity-flip transitions generates a non-slash
term in the quark propagator, i.e. the dynamically-generated mass
$M(p)$, see Fig. 4. It implies the spontaneous chiral symmetry breaking.

\begin{figure}[t]
\epsfig{file=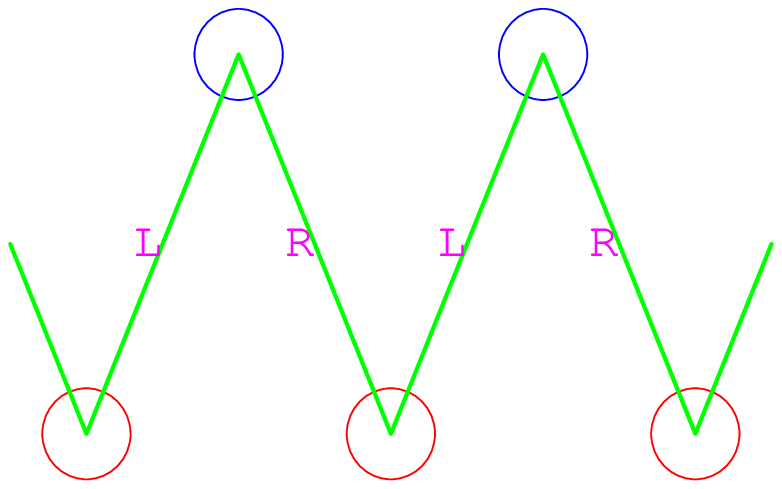,width=5.5cm}
\hspace{-.01cm}\epsfig{file=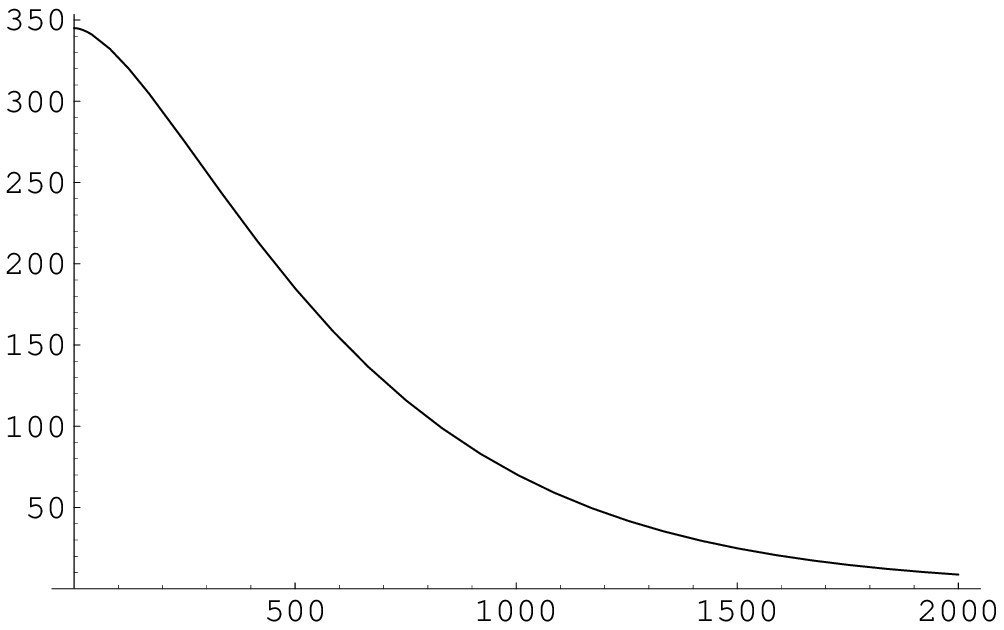,width=6cm}
\vspace{-.01cm}\caption{Quarks hopping from instantons to anti-instantons
and {\it vice versa} flip helicity (left). An infinite number
of such jumps generates a dynamical mass $M(p)$, in ${\rm MeV}$
(right).~$^3$
\label{propag}}
\end{figure}

Two different formalisms have been developed
in the 80's to calculate hadron observables: ({\it i}) first computing
an observable and then averaging it over the instanton ensemble \cite{DP1}
and ({\it ii}) first averaging over the ensemble which leads to an
effective low-energy theory, and then computing observables in the effective
theory. \cite{DP2} Despite very different appearance the two formalisms give
identical final results. Let me list a few of them:
\beq
<\bar q q>=-\frac{{\rm const.}}{\bar R^2\bar\rho}\simeq -(255\,{\rm
MeV})^3,\qquad
M(0)={\rm const.}\,\frac{\pi \bar\rho}{\bar R^2}\simeq 345\,{\rm MeV},
\la{r1}\eeq
\beq
F_\pi={\rm const.}\,\frac{\bar\rho}{\bar R^2}
\sqrt{\log\frac{\bar R}{\bar\rho}}\simeq 100\,{\rm MeV}\quad{\rm vs.}\quad
94\,{\rm MeV\,(exper)},
\la{r2}\eeq
\beq
m_{\eta^\prime}=\frac{{\rm const.}}{\bar\rho}\simeq 980\,{\rm MeV}
\quad{\rm vs.}\quad 958\,{\rm MeV\,(exper)}\ldots
\la{r3}\eeq
where ``${\rm const.}$'' are computable numerical constants of the
order of unity.

Recently, the instanton mechanism of the SCSB has been scrutinized by
direct lattice methods. \cite{N2,Da,Net} At present there is one group
\cite{Net} challenging the instanton mechanism. However, the density
of alternative `local structures' found there explodes as the lattice
spacing decreases, and this must be sorted out first. Studies by other
groups \cite{N2,Da} support or strongly support the mechanism described
above. In particular, in a recent paper Gattringer~\cite{Da}
convincingly demonstrates that quarks in near-zero modes concentrate
in the regions where the gluon field is either self-dual ($I$'s) or
anti-self-dual ($\bar I$'s). Since near-zero modes are responsible
for the SCSB it is a direct confirmation of the instanton mechanism. 

\section{Baryons}
There is a remarkable evidence of the importance of
instantons for the baryon structure. In Ref. \cite{N} the so-called
density-density correlation function inside the nucleon has been measured
both in the full vacuum and in the instanton vacuum resulting from the full
one by means of smoothing. The correlation in question is between the
densities of $u$ and $d$ quarks separated by a distance $x$ inside the
nucleon which is created at some time and annihilated at a later time.
The two correlators (`full' and `instanton') are depicted in Fig. 5: one
observes a remarkable agreement between the two, up to $x=1.7\,{\rm fm}$.

\begin{figure}[t]
\epsfig{file=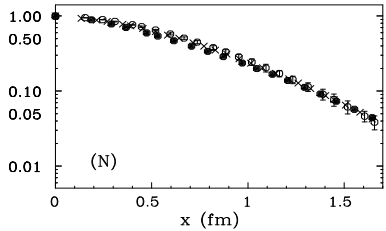,height=3cm,width=5cm}
\vspace{-.01cm}\caption{Density-density correlation function
in the nucleon. $^{11}$ Filled circles are measurements in the
full gluon vacuum (corresponding to Fig. 3a,b) while open circles
are measured in the vacuum with instantons only (Fig. 3c,d).
Despite that linear confining potential is absent in the instanton
vacuum the nucleon structure seems to be very well reproduced.
\label{Ncorr}}
\end{figure}

It must be stressed that neither the one-gluon exchange nor the
linear confining potential present in the full gluon vacuum
survive the smoothing of the gluon field shown in Fig. 3.
Nevertheless, quark correlations in the nucleon remain basically
unaltered! It means that {\bf neither the one-gluon exchange nor the
linear confining potential are important for the quark binding
inside the nucleon}. As a matter of fact, the same remark can be
addressed to the lightest mesons $\pi$ and $\rho$ since the
density-density correlators for these hadrons also remain basically
unchanged as one goes from the full glue to the reduced instanton
vacuum. \cite{N} Therefore, one must be able to explain at least
the lightest $\pi,\rho,N$ on the basis of instantons only.

The dynamics remaining in the instanton vacuum is the SCSB, the
apearance of the dynamical quark mass $M(p)$, and quark interactions
induced by the possibility that they scatter off the same instanton.
Actualy these interactions named after 't Hooft, are quite strong.
\cite{DP2,DSS} They are in fact so strong that for quark and antiquark
in the pion channel they eat up the $700\,{\rm MeV}$ of twice the
constituent quark mass to nil, as required by the Goldstone theorem.
In the vector meson channel 't Hooft interactions are suppressed,
and that is why the $\rho$ mass is roughly twice the constituent quark
mass. In the nucleon they are fully at work but in a rather peculiar way:
instanton-induced interactions can be iterated as many times as one
wishes in the exchanges between quarks, see Fig. 6, left. It can be easily
verified that the diagram in Fig. 6, left, can be drawn as three continuous
quark lines going from the l.h.s of the diagram to its r.h.s., without
adding closed loops. Therefore, that kind of interaction arises
already in the so-called quenched approximation. At the same time, it
yields plenty of Z-graphs absent in ``valence QCD'' but which are
necessary to reproduce hadron properties. \cite{Liu} 

%%%%%%%%%%%%%%%%%%%%%%%%%%
% FIGURE 6               %
%%%%%%%%%%%%%%%%%%%%%%%%%%

\begin{figure}[t]
\hspace{2cm}\epsfig{file=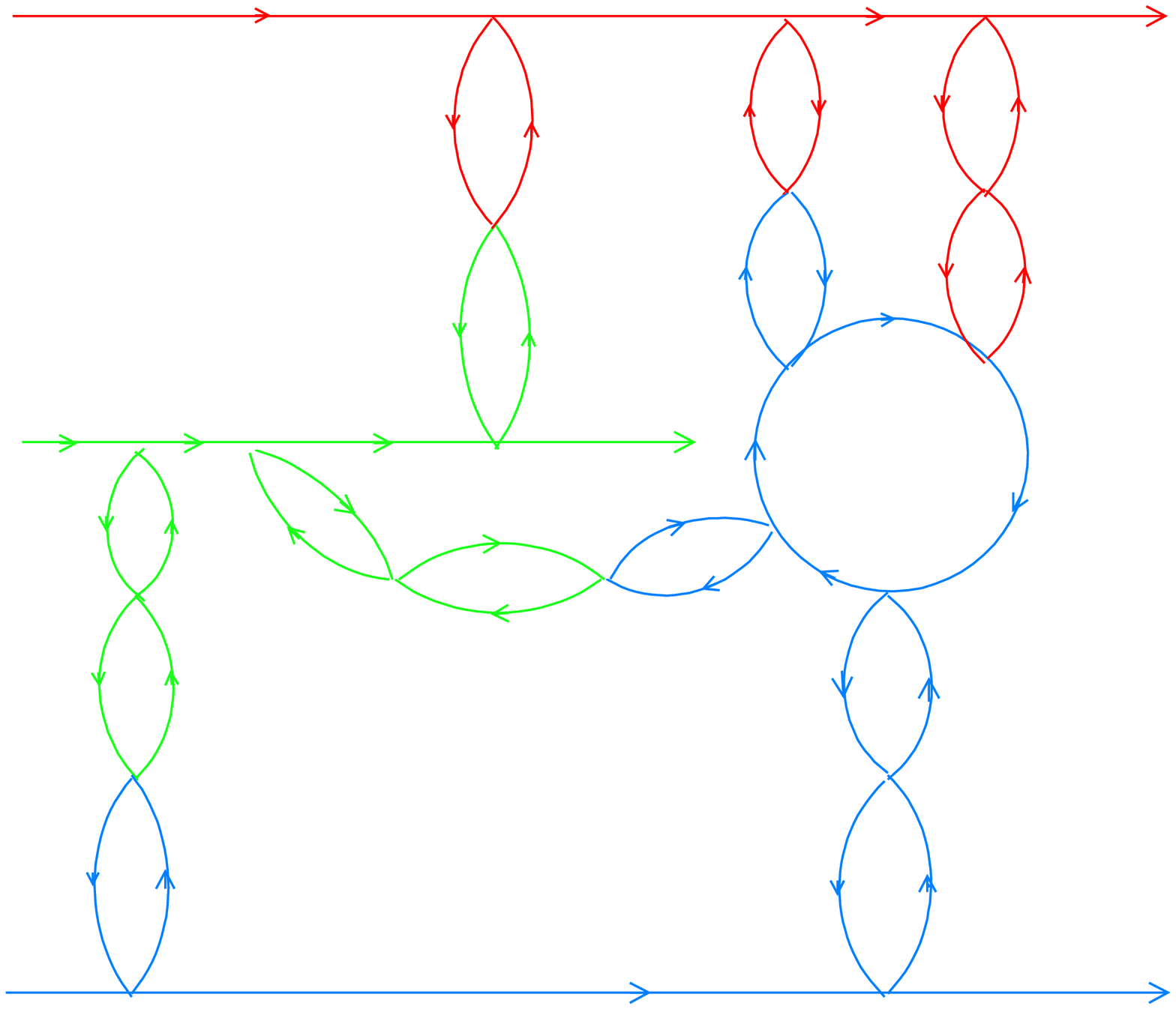,height=3.9cm,width=4.5cm}
%\vskip -2.2true cm
\hspace{.5cm}
\epsfig{file=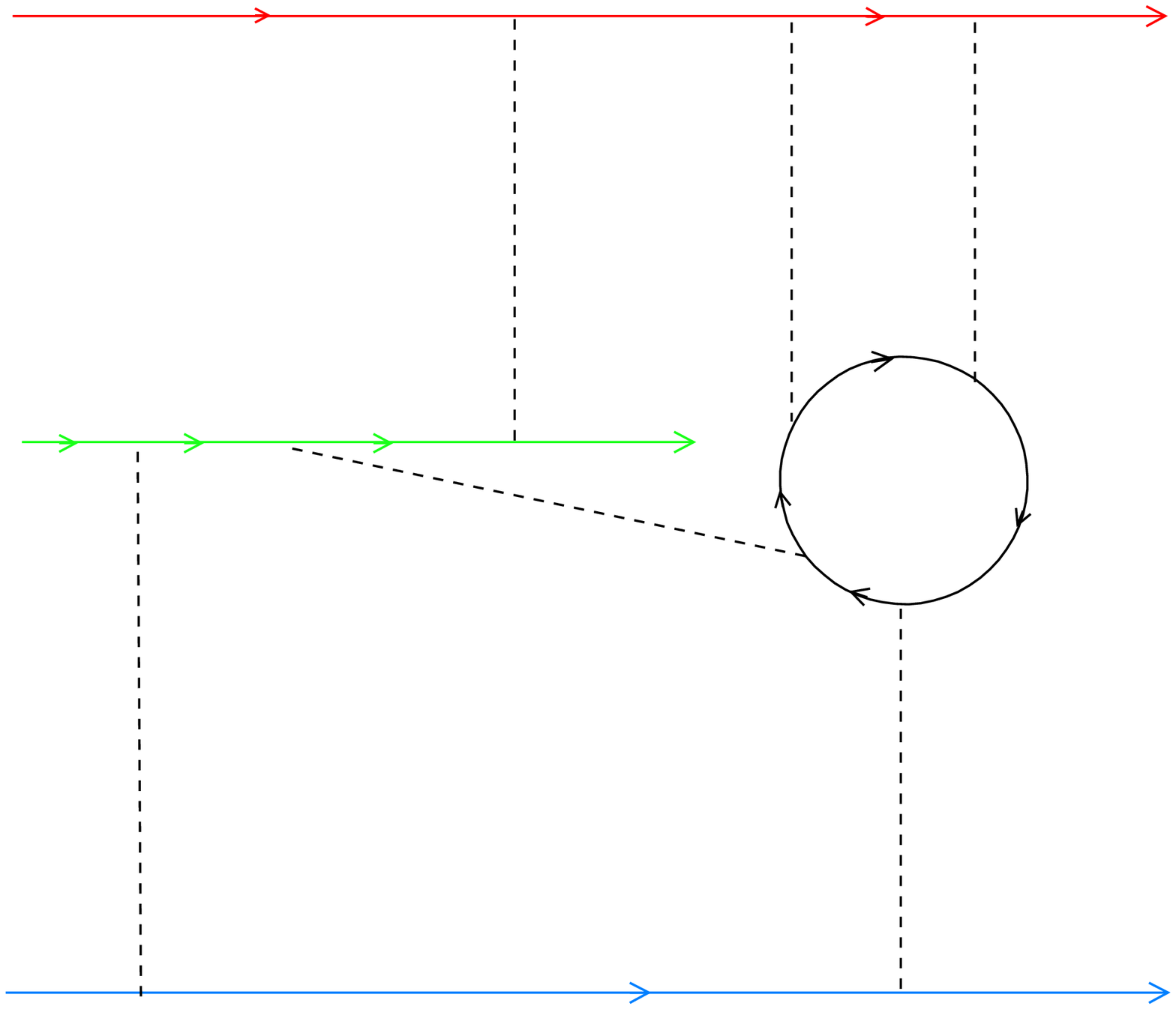,height=3.9cm,width=4.5cm}
\vspace{-.8cm}
\caption{'t Hooft interactions in the nucleon (left)
essentially come to quarks interacting via pion fields (right).
\label{nucl}}
\end{figure}

Summing up all interactions of the kind shown in Fig. 6, left, seems
to be a hopeless task. Nevertheless, the nucleon binding problem can be
solved {\em exactly} when two simplifications are used. The first exploits
the fact that in the instanton vacuum there are two lightest degrees
of freedom: pions (since they are the Goldstone bosons) and quarks
with the dynamical mass $M$. All the rest collective excitations of
the instanton vacuum are much heavier, and one may wish to neglect them.
Pions arise from summing up the $q\bar q$ bubbles schematically shown
in Fig. 6, left. The resulting effective low-energy theory takes the
form of the non-linear $\sigma$-model: \cite{DP2,Penisc}
\beq
{\cal L}_{\rm eff}=\bar q\,
\left[i\dd-M\exp(i\gamma_5\pi^A\tau^A/F_\pi)\right]\,q.
\la{d2}\eeq
The absence of the explicit kinetic energy term for pions (which
would lead to the double counting) distinguishes it from the
Manohar--Georgi model. \cite{MG} Expanding the exponent to the
first power in $\pi^A$ we find that the dimensionless pion--constituent
quark coupling,
\beq
g_{\pi qq}=\frac{M}{F_\pi}\approx 4,
\la{coupl}\eeq
is quite strong. The domain of applicability of the low-energy
effective theory \ur{d2} is restricted by momenta
$p<1/\bar\rho=600\,{\rm MeV}$, which is the inverse size of constituent
quarks. At higher momenta one starts to feel the internal structure of
constituent quarks, and the two lightest degrees of freedom of \Eq{d2}
become insufficient. However, the expected typical momenta of quarks in
the nucleon are of the order of $M\approx 345\;{\rm MeV}$, which is inside
the domain of applicability of the low-momentum effective theory.

The chiral interactions of constituent quarks in the nucleon,
following from the effective theory \ur{d2}, are schematically shown
in Fig. 6, right, where quarks are denoted lines with arrows. Notice that,
since there is no explicit kinetic energy for pions in \Eq{d2}, the pion
propagates only through quark loops. Quark loops induce also many-quark
interactions indicated in Fig. 6 as well. We see that the emerging picture is
rather far from a simple one-pion exchange between the constituent quarks: the
non-linear effects in the pion field are essential.

The second simplification is achieved in the limit
of large $N_c$. For $N_c$ colors the number of constituent quarks in a
baryon is $N_c$ and all quark loop contributions are also proportional
to $N_c$. Therefore, at large $N_c$ one can speak about a
{\em classical self-consistent pion field} inside the nucleon:
quantum fluctuations about the classical field are suppressed as $1/N_c$.
The problem of summing up all diagrams of the type shown in Fig. 6
is thus reduced to finding a classical pion field pulling $N_c$ massive
quarks together to form a bound state.

\section{Chiral Quark--Soliton Model}
Let us imagine  a classical time-independent pion
field which is strong and spatially wide enough to form a bound-state
level in the Dirac equation following from \Eq{d2}. The background
chiral field is color-neutral, so one can put $N_c$ quarks
on the same level in an antisymmetric state in color, i.e.
in a color-singlet state. Thus we obtain a baryon state, as compared
to the vacuum.

One has to pay for the creation of this trial pion field, however.
Since there are no terms depending directly on the pion field in the
low-momentum theory \ur{d2} the energy of the pion field is actually
encoded in the shift of the lower negative-energy Dirac sea of quarks,
as compared to the free case with zero pion field. The baryon mass
is the sum of the bound-state energy and of the aggregate energy
of the lower Dirac sea. It is a functional of the trial pion field;
one has to minimize it with respect to that field to find the
self-consistent pion field that binds quarks inside a baryon. It is
a clean-cut problem, and can be solved numerically or, approximately,
analylically.  The description of baryons based on this construction
has been named the Chiral Quark--Soliton Model (CQSM). \cite{KRS,BB,DP5}

The model reminds the large-$Z$ Thomas--Fermi atom where
$N_c$ plays the role of $Z$. Fortunately, corrections to the model go as
$1/N_c$ or even as $1/N_c^2$ and have been computed for many observables.
In the Thomas--Fermi model of atoms corrections to the self-consistent
(electric) field are of the order of $1/\sqrt{Z}$ and for that reason
are large unless atoms are very heavy.

In the end of the 80's and the beginning of the 90's dozens of baryon
characteristics have been computed in the CQSM, including masses,
magnetic moments, axial constants, formfactors, splittings inside the
mutliplets and between multiplets, polarizability, fraction of nucleon
spin carried by quarks, etc. --
see \cite{Review,DP3} for a review and references therein. Starting from '96 a
new class  of problems have been addressed, namely parton distributions in the
nucleon at low virtuality. \cite{SF} Parton distributions are a snapshot of
the nucleon in the infinite momentum frame. One needs an inherently
relativistic model in order to describe them consistently. For
example, a bag model or any other nonrelativistic model with three quarks in a
bound state, being naively boosted to the infinite-momentum frame gives a
{\it negative} distribution of antiquarks, which is nonsense. On the contrary,
being a relativistic field-theoretic model CQSM predicts parton distributions
that satisfy all general requirements known in full QCD, like positivity and
sum rules constraints.

Numerous parton distributions have been computed in the CQSM, mainly
by the Bochum group. \cite{SF,PPGWW,DGPW} There have been a number of
mysteries from naive quark models' point of view: the large number of
antiquarks already at a low virtuality, the `spin crisis', the large flavor
asymmetry of {\it anti}quarks, etc. The CQSM explains all those `mysteries'
in a natural way as it incorporates, together with valence quarks bound by the
isospin-1 pion field, the negative-energy Dirac sea. Furthermore, the CQSM
predicts nontrivial phenomena that have not been observed so far: large
flavor asymmetry of the {\em polarized} antiquarks \cite{DGPW}, transversity
dictributions \cite{PP},  peculiar shapes of the so-called skewed parton
distributions \cite{PPPBGW} and other phenomena in hard exclusive
reactions.~\cite{GPV} Baryon dynamics is rich and far from naive
``three quarks'' expectations.

\section{Conclusions}

\hskip .7true cm
1. The would-be linear confining potential of the pure glue
world is necessarily screened by pion production at very moderate separations
between quarks. Therefore, light hadrons should not be sensitive to
confinement forces but rather to the dynamics of the spontaneous chiral
symmetry breaking (SCSB).

2. Most likely, the SCSB is driven by instantons -- large nonperturbative
fluctuations of the gluon field having the meaning of tunneling. The SCSB is
due to `hopping' of quarks from one randomly situated instanton to another,
each time flipping the helicity. The instanton theory of the SCSB is in
agreement with the low-energy phenomenology
({\it cf.} the chiral condensate
$<\bar q q>$, the dynamical quark mass $M(p)$, $F_\pi$, $m_{\eta^\prime}$...)
and seems to be confirmed by direct lattice methods.
Furthermore, lattice simulations indicate that instantons alone are
responsible for the properties of lightest hadrons $\pi,\rho,N,...$

3. Summing up instanton-induced quark interactions in baryons
leads to the Chiral Quark--Soliton Model where baryons appear to
be bound states of constituent quarks pulled together by the
chiral field. The model enables one to compute numerous parton distributions,
as well as `static' characteristics of baryons -- with no fitting parameters.

4. For highly excited baryons ($m\!=\!1.5\!-\!3\,{\rm GeV}$) the relative
importance of confining forces {\it vs.} those of the SCSB may be reversed.
One can view a large-spin $J$ resonance as due to a short-time stretch of an
unstable string or, alternatively, as a rotating elongated pion cloud.
\cite{DP6} What picture is more adequate is a question to experiment. In the
first case the dominant decay is on the average of the type ${\rm Bar}_{J}\to
{\rm Bar}_{\sim J/2} +{\rm Mes}_{\sim J/2}$; in the second case it is
mainly a cascade ${\rm Bar}_{J}\to {\rm Bar}_{J-1}+\pi
\to {\rm Bar}_{J-2}+\pi\pi\to ...$ Studying resonances can
elucidate the relation between chiral and confining forces.

\end{document}